\documentclass[12pt]{article}
\usepackage{epsfig,a4wide}

\usepackage{amsmath}
\usepackage{amssymb}
\usepackage{graphicx}

\newcommand{\be}{\begin{equation}}
\newcommand{\ee}{\end{equation}}
\newcommand{\ba}{\begin{eqnarray}}
\newcommand{\ea}{\end{eqnarray}}

\begin{document}

\title{Study of possible $\omega$ bound states in nuclei with 
the $(\gamma , p)$ reaction}
\author{M. Kaskulov$^a$,  H. Nagahiro$^b$, S. Hirenzaki$^c$, E. Oset$^a$ \\
{$^a$\small Departamento de F\'{\i}sica Te\'orica and IFIC,
Centro Mixto Universidad de Valencia-CSIC,} \\ 
{ \small  Institutos de
Investigaci\'on de Paterna, Aptd. 22085, 46071 Valencia, Spain} \\
{$^b$\small Research Center for Nuclear Physics, Osaka University, Ibaraki, Osaka
567-0047 Japan}\\
{$^c$\small Department of Physics, Nara Women's University, Nara 630-8506
Japan}\\
}

\date{\today}
\maketitle 
\begin{abstract}  

We perform calculations for $\omega$ production in nuclei by means of the
$(\gamma , p )$ reaction for photon energies and proton angles suited to
running and future experiments in present Laboratories.  For some cases of
possible $\omega$ optical potentials we find clear peaks which could be
observable provided a good resolution in the $\omega$ energy is available. We
also study the inclusive production of $\pi ^0 \gamma$ in nuclei around the 
$\omega$ energy and find a double hump structure for the energy spectra, 
 with a peak around a $\pi ^0 \gamma$ energy of 
$m_{\omega}-100$~MeV, which could easily be misidentified by a signal of
  a bound $\omega $ state in nuclei, while it is due to a different scaling of the 
  uncorrelated $\pi ^0 \gamma$ production and $\omega$ production with
  subsequent $\pi ^0 \gamma$ decay. 

\end{abstract}

\section{Introduction} 

The interaction of hadrons with nuclei is one of the important chapters in
hadron and nuclear physics and much work has been devoted to it
\cite{Post:2003hu}. In particular the behavior of vector mesons
in nuclei has received much attention,
stimulated by the ansatz of a universal scaling of the vector meson masses in
nuclei suggested in \cite{scaling} and the study of QCD sum rules  in
nuclei \cite{hatsuda},
although earlier studies within the Nambu-Jona-Lasinio model produced no
dropping of the vector meson masses \cite{bernard}. 
 More concretely, the properties of the 
$\omega$ meson have been thoroughly studied theoretically and
 different calculations have been carried out
within varied models ranging from quark models, to phenomenological 
evaluations,
or using effective Lagrangians \cite{kurasawa,jean,klingl1,saito,tsushima,
friman,klingl2, post, saito2,lykasov,sibirtsev,dutt,lutz,zschocke,dutt2,
muhlich,mosel,muhlich2,steinmueller}. 
 The values obtained for the selfenergy of 
the $\omega$ in nuclei split nearly equally into attracion and repulsion 
and range from an attraction of the order of 100-200 MeV 
\cite{tsushima,klingl2} to no changes in the mass \cite{mosel} to a net
repulsion of the order of 50 MeV \cite{lutz}.

Experimental work along these lines is also rich and recently the NA60
collaboration \cite{Arnaldi:2006jq} has produced dilepton spectra of 
excellent mass resolution in
heavy ion reactions, for the spectra of the $\rho$, which points to a large
broadening of the $\rho$ but no shift on the mass.

  On the other hand, it has been argued in \cite{moseltalk} that reactions
involving the interaction of elementary particles with nuclei can be equally
good to show medium effects of particles, with the advantage of being easier to
analyse. In this sense, a variety of experiments have been done with $pA$
collisions in nuclei at KEK \cite{ozawa,tabaru,sakuma} and photonuclear collisions at
Jefferson lab \cite{weygand} by looking at dilepton spectra. 

   A different approach has been followed by the CBELSA/TAPS collaboration by
   looking at the $\gamma \pi ^0$ coming from the $\omega$ decay. 
In this line a    recent work \cite{trnka}
claims evidence for a decrease of the $\omega$ mass in the medium of the
order of 100 MeV from the study of the modification of the mass
spectra in $\omega$ photoproduction ( actually, the conclusions of this paper
are tied to the way the background is subtracted and it will be shown in
\cite{murat} that with other justified choices of background there 
is no shift of the mass or it could even be positive).

   With sufficient attraction, $\omega$ bound states could be produced, and
could be even observable provided the $\omega$ width in the medium would not be
too large.  Indeed, several works have investigated the possibility to have
$\omega$ bound states in nuclei \cite{marco,tsushima} and speculations on this
possibility are also exploited in \cite{trnkathesis}.  Suggestions to measure
such possible states with the (d,He-3) recoilless reaction have also been made
\cite{EPJA6}.  

In the present work, and stimulated by the work of 
\cite{trnkathesis}, we shall study the photon induced $\omega$ production in
nuclei, looking both at the experimental set up of \cite{trnkathesis}, as well
as to other set ups which we consider more suited to see bound $\omega$ states 
with this reaction. We will make predictions for cross sections for 
the $(\gamma, p)$ reaction in nuclei, leading to $\omega$ bound states, 
for several photon energies and proton angles . 

 At the same time we shall also 
present results for inclusive $\omega$ production, looking at the 
$\gamma \pi ^0$ decay mode of the $\omega$, as in \cite{trnkathesis}, and will
show that due to the presence of an unavoidable background of $\gamma \pi ^0$
(unrelated to the $\omega$) at $\gamma \pi ^0$ energies smaller than the $\omega$
mass, and to the different A-mass dependence of the background and $\omega$
production, a peak develops around $m_{\omega} - 100~ MeV$ in nuclei, 
which we warn not to
misidentify with a signal of a bound $\omega$ state in the nucleus.  We shall
also show the optimal condictions to observe signals of eventual $\omega$ bound 
states, as well as the minimal experimental resolution necessary to 
see the possible peaks.

\section{Production of bound $\omega$ states in the ($\gamma$,p)  
reaction}

Here we evaluate the formation rate of  $\omega$ bound states in the 
nucleus by means of the ($\gamma$,p) reaction.  We use the  
Green function method \cite{NPA435etc} to calculate the cross  
sections for $\omega$-mesic states formation as described in
Refs.~\cite{NPA761,nuclth0606052} in   
detail.
The theoretical model used here is exactly same as that used in these
references.

\begin{figure}[hbt]
\center
\includegraphics[width=9cm]{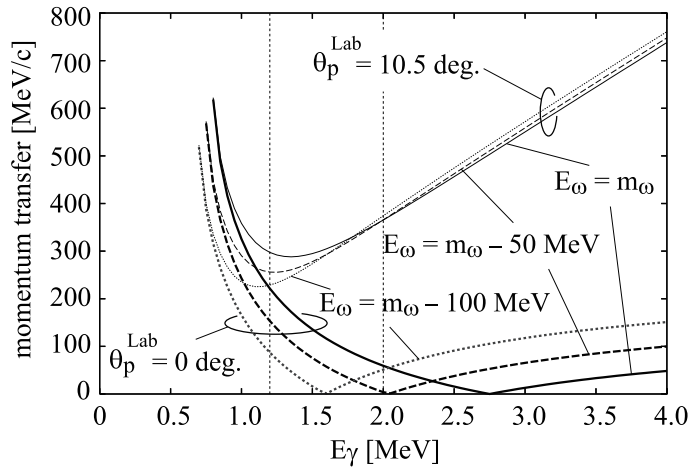}
\caption{\small
Momentum transfers are shown as a function of the incident photon energy
 $E_\gamma$ in the ($\gamma$,p) reaction. The solid, dashed and dotted
 lines show the 
 momentum transfers at $\omega$ energy $E_\omega = m_\omega$,
$E_\omega = m_\omega - 50$ MeV and $E_\omega = m_\omega -100$ MeV,
 respectively. The thick lines indicate the forward reaction cases and
 the thin lines 
 the cases for the ejected proton in the final state with the finite angle
$\theta_p^{\rm Lab} = 10.5$ degree.
The vertical dashed lines show the incident energies
$E_\gamma=1.2$ GeV and 
$2.0$ GeV.
\label{fig:mom_trans}
}
\end{figure}

We show first the momentum transfer of the ($\gamma$,p) reaction in  
Fig.~\ref{fig:mom_trans}, as a function of the incident photon energy, at
forward and finite angles of the emitted proton.
The momentum transfer is the important kinematical variable which determines the
experimental feasibility of the formation of meson-nuclear bound states.
Indeed, deeply bound pionic
atoms were discovered experimentally at the recoilless kinematics
\cite{NPA530,PLB527,EPJA6,PRC62etc}. 
In the formation of $\omega$ states 
we find that the
recoilless condition is satisfied at $E_\gamma \sim 2.7$ GeV, as in 
Ref.~\cite{marco},
for 
 $\omega$ production at threshold  and
 forward proton production. However, the recoilless condition is never  
satisfied for finite proton angles. Since some experimental set ups can have 
problems in proton forward detection, it is interesting to
determine the optimal conditions with these boundary conditions. For this
purpose we look at the optimal photon energy for protons emitted with a finite
angle. For an angle of 
$\theta_p^{\rm lab}=10.5$  
degrees for the emitted proton, the angle measured in  \cite{trnkathesis}, 
the momentum transfer takes the minimum value at $E_\gamma  
\sim 1.2$ GeV.  In this section, we consider  1.2 and 2.0 GeV  
as the incident photon energies and 0 and 10.5 degrees as the emitted  
proton angles in the laboratory frame.

In order to calculate the cross sections at finite angles of the 
emitted proton, 
we estimate the elementary cross sections from the  
experimental data shown in Tables 3--5 in Ref.~\cite{EPJA18},
and we use 5.0 $\mu$b/sr ($\theta_p^{\rm Lab}=0$ deg.) and 8.0 $\mu$b/sr
(10.5 deg.) at   
$E_\gamma= 1.2$ GeV, and 0.7 $\mu$b/sr for both $\theta_p^{\rm Lab}=0$
and 10.5 deg.   
at $E_\gamma=2.0$ GeV in the laboratory frame, respectively.

The $\omega$-nucleus optical potential is written here as;
\begin{equation}
V(r) = (V_0 + iW_0  ) \frac{\rho(r)}{\rho_0},
\end{equation}
where  $\rho(r)$ is the nuclear experimental density for which we take 
 the two parameter Fermi  distribution.
We consider three cases of the potential strength as;
\begin{subequations}
\begin{eqnarray}
(V_0,W_0) &=& -(0,50) \label{eq:0_50}\\
&=&-(100,50) \label{eq:100_50}\\
&=&-(156,29) \label{eq:156_29}
\end{eqnarray}
\end{subequations}
in unit of MeV. The reason for these choices is as follows. From
\cite{trnkathesis} on the  $\omega$ production rates in different nuclei one
deduces a width for the $\omega$ at the average $\omega$ momentum in the
production ( $\sim 900 MeV$) and $\rho = \rho _0$ of about $ 100 ~ MeV$
\cite{cola}.  This means that the imaginary part of the potential has a strength
of about $50 ~ MeV$. As discussed above, uncertainties in the
subtraction of background in the  experiment of \cite{trnka} do not allow us to
draw any conclusion on the shift of the mass \cite{murat}.  Thus we have kept 
the possibility of a downward shift of the mass open  and have performed 
calculations for 100 MeV binding too. We also consider the potential
estimated theoretically shown in Eq.~(\ref{eq:156_29}), which is
obtained by the linear density approximation with the scattering length
$a=1.6 + 0.3 i$ fm \cite{klingl2}.
This potential in Eq.~(\ref{eq:156_29}) is strongly attractive with weak
absorption and hence should be the ideal case for the formation
of $\omega$ mesic nuclei.


\begin{figure}[hbt]
\center
\includegraphics[width=9cm]{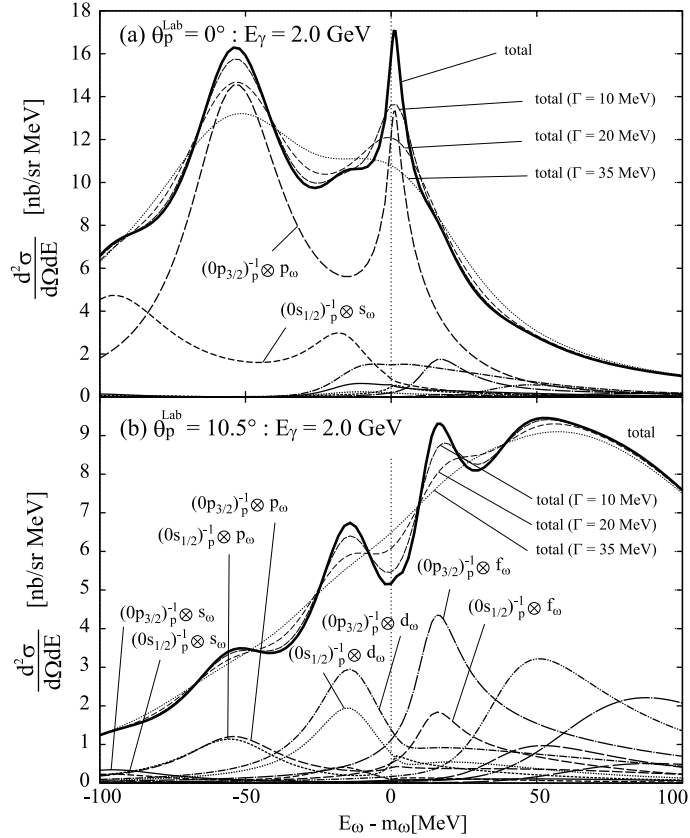}
\caption{\small
 Formation spectra of the $\omega$ mesic nucleus in
 $^{12}$C($\gamma$,p) reaction at emitted proton angle (a)
 $\theta_p^{\rm Lab}=0$ degree and (b) $\theta_p^{\rm Lab}=10.5$ degree
 calculated with the potential depth $(V_0,W_0) = -(156,29)$ MeV
 as in Eq.~(\ref{eq:156_29}).
The incident photon energy is $E_\gamma=2.0$ GeV.
The thick solid lines show the total spectra and the dashed lines the
 subcomponents as indicated in the figures. The assumed experimental
 resolutions are also indicated in the figures.
\label{fig:2.0_156_29}
}
\end{figure}


\begin{figure}[hbt]
\center
\includegraphics[width=9cm]{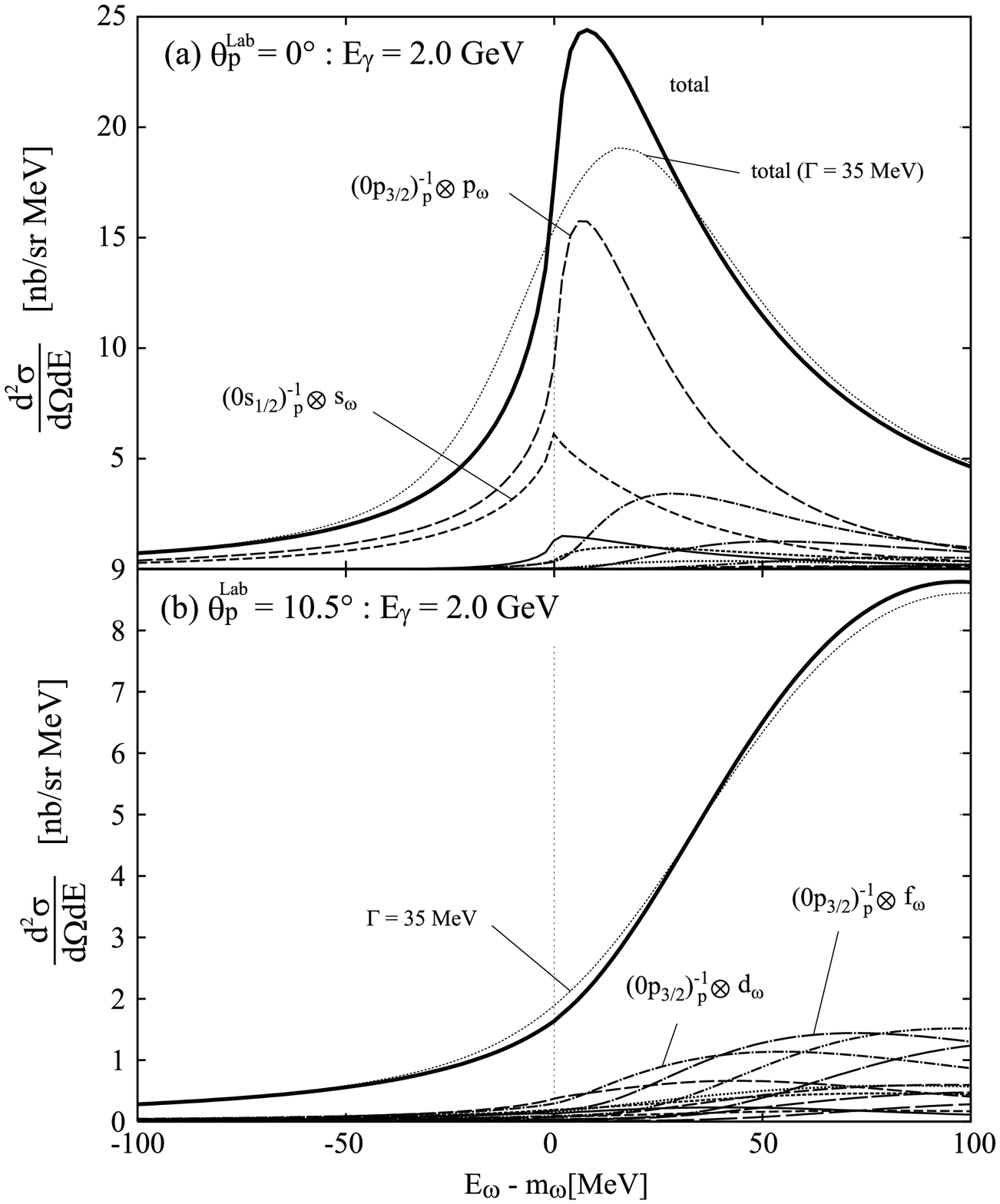}
\caption{\small
Same as Fig.~\ref{fig:2.0_156_29} except for the potential depth
$(V_0,W_0) = -(0,50)$ MeV 
as in Eq.~(\ref{eq:0_50}).
\label{fig:2.0_0_50}
}
\end{figure}
\begin{figure}[hbt]
\center
\includegraphics[width=9cm]{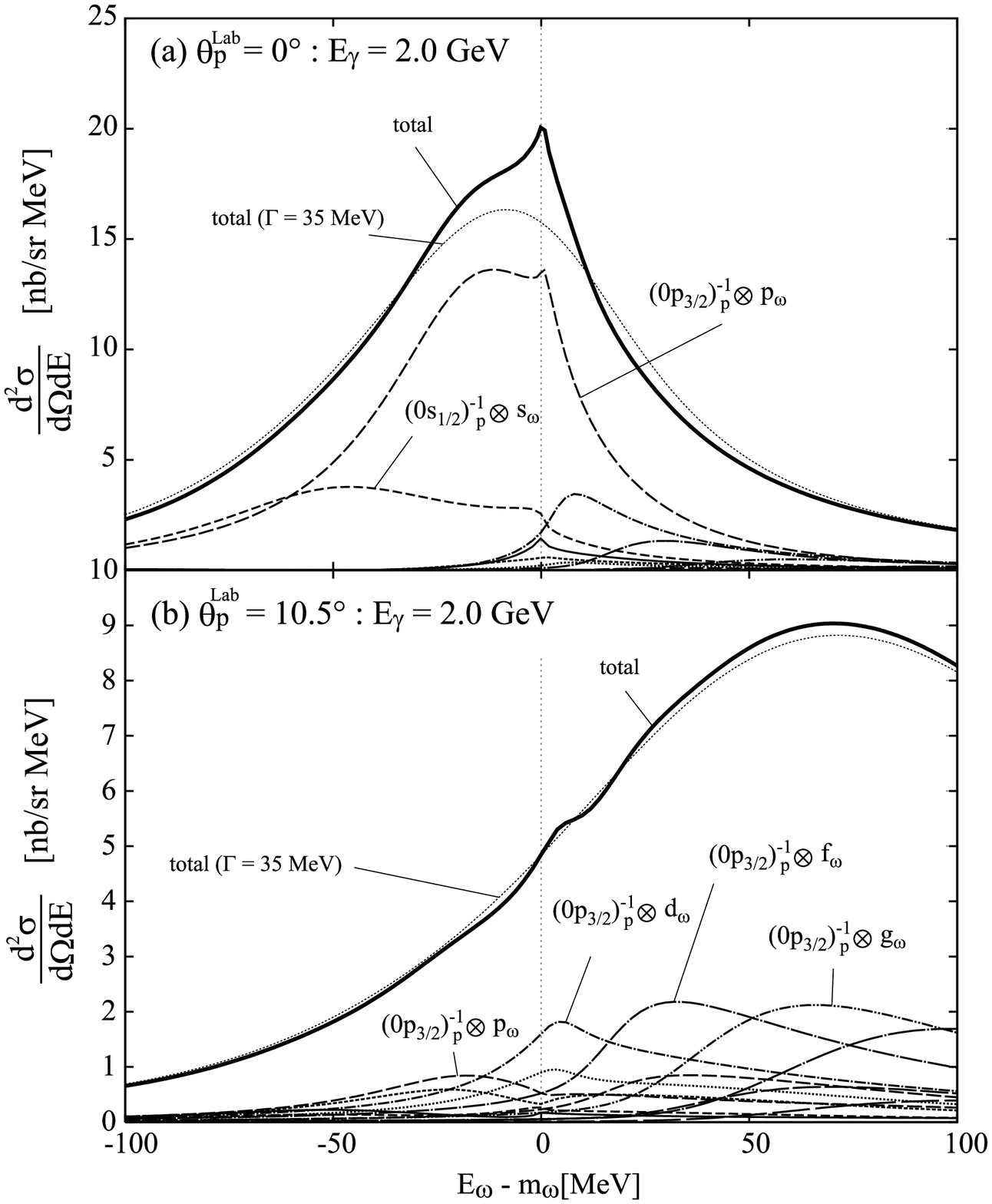}
\caption{\small
Same as Fig.~\ref{fig:2.0_156_29} except for the potential depth
$(V_0,W_0) = -(100,50)$ MeV 
as in Eq.~(\ref{eq:100_50}).
\label{fig:2.0_100_50}
}
\end{figure}

\begin{figure}[hbt]
\center
\includegraphics[width=9cm]{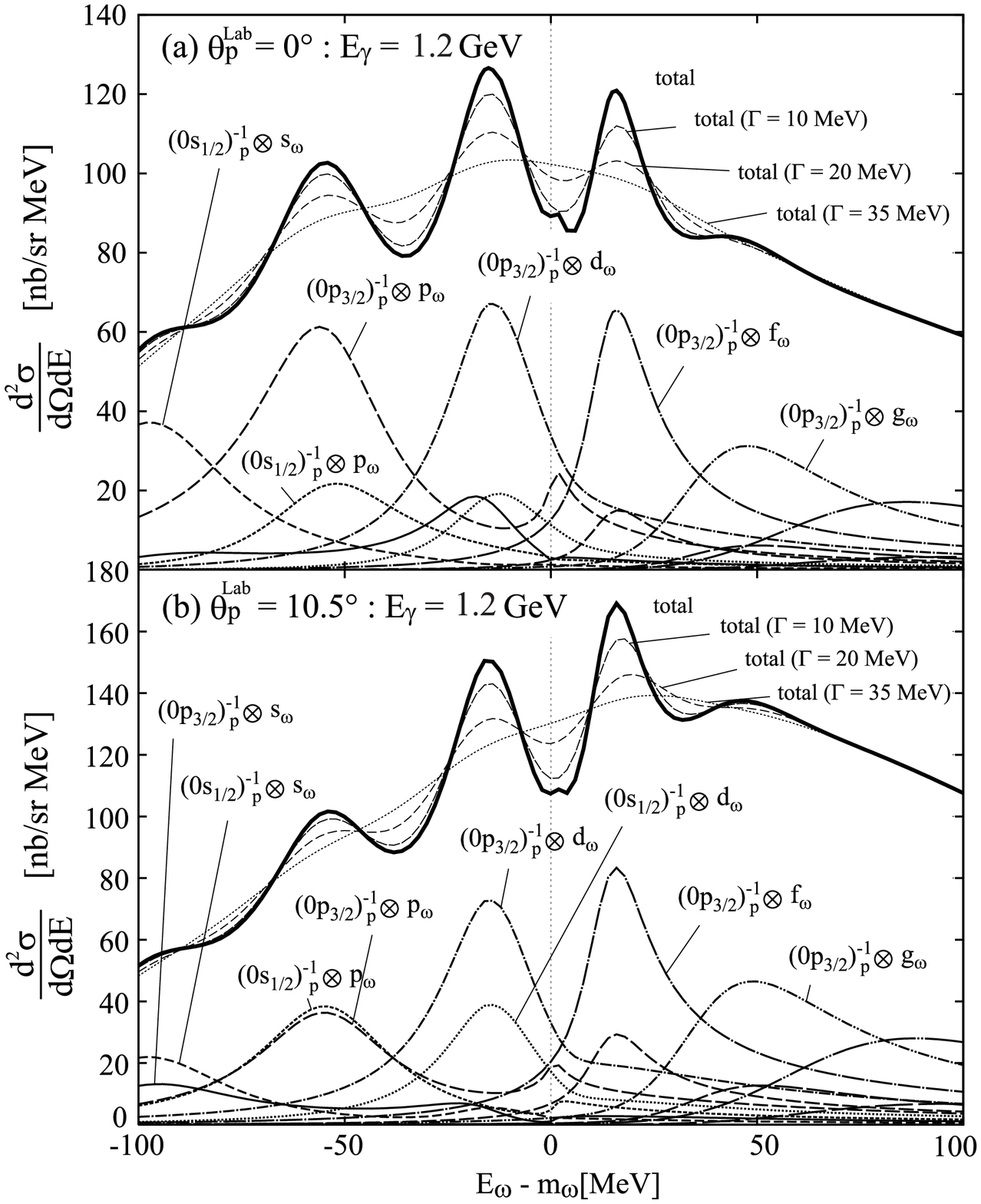}
\caption{ \small
 Formation spectra of the $\omega$ mesic nucleus in
 $^{12}$C($\gamma$,p) reaction at emitted proton angle (a)
 $\theta_p^{\rm Lab}=0$ degrees and (b) $\theta_p^{\rm Lab}=10.5$ degrees
 calculated with the potential depth $(V_0,W_0) = -(156,29)$ MeV
 as in Eq.~(\ref{eq:156_29}).
The incident photon energy is $E_\gamma=1.2$ GeV.
The thick solid lines show the total spectra and the dashed lines the
 subcomponents as indicated in the figures. The assumed experimental
 resolutions are also indicated in the figures.
\label{fig:1.2_156_29}
}
\end{figure}
\begin{figure}[hbt]
\center
\includegraphics[width=9cm]{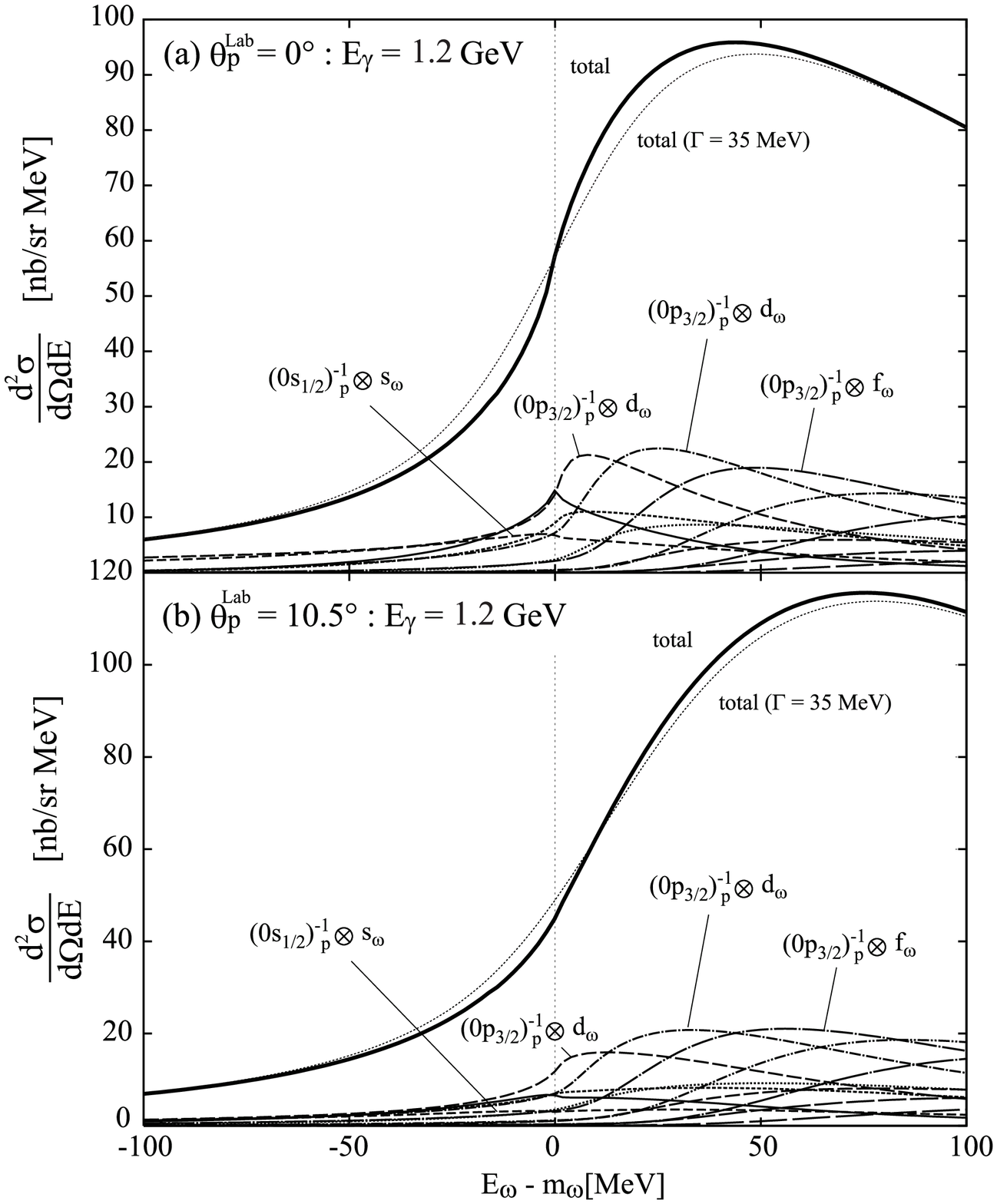}
\caption{\small
Same as Fig.~\ref{fig:1.2_156_29} except for the potential depth
$(V_0,W_0) = -(0,50)$ MeV 
as in Eq.~(\ref{eq:0_50}).
\label{fig:1.2_0_50}
}
\end{figure}
\begin{figure}[hbt]
\center
\includegraphics[width=9cm]{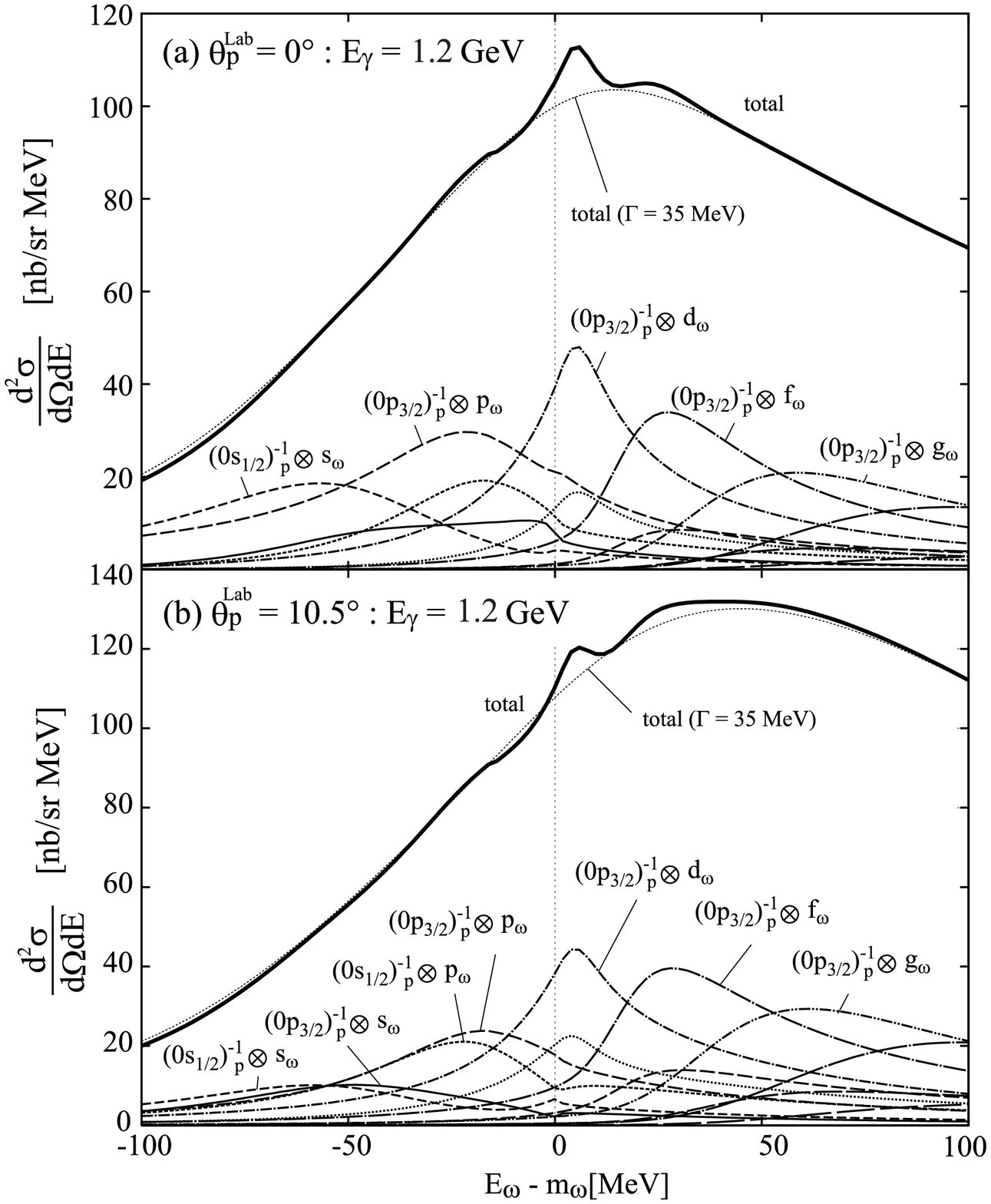}
\caption{
\small
Same as Fig.~\ref{fig:1.2_156_29} except for the potential depth
$(V_0,W_0) = -(100,50)$ MeV 
as in Eq.~(\ref{eq:100_50}).
\label{fig:1.2_100_50}
}
\end{figure}

No $\omega$ bound states are expected for the potential in Eq.~(\ref{eq:0_50}) 
which has only an absorptive part.
The potential in Eq.~(\ref{eq:100_50}) has a strong attraction 
with the large absorptive part as indicated in
Ref.~\cite{trnkathesis}.
It is also interesting to compare the  
formation spectra obtained with the potentials in
Eq.~(\ref{eq:100_50}) and (\ref{eq:156_29}) to know the effects of the
$\omega$ absorption in nuclei  

We should mention here that the all spectra in this section are  
plotted as functions of $E_\omega - m_\omega$, while
in previous papers \cite{NPA761,nuclth0606052,NPA530,PLB527,EPJA6} they were
plotted as functions   
of excitation energies of final mesic-nuclear state, 
or equivalently, the energies of emitted particles
which included the excitation energies of the core nucleus, too.  We  
plot the spectra in this manner since  
we assume here experiments in which the
$\omega$ meson energy can be deduced separately from the nuclear core
excitation. This is the case here where the $\omega$ energy is measured  
by the $\pi^0$ and  
$\gamma$ observation
from subsequent $\omega$ decay 
$\omega \rightarrow \pi^0 \gamma$ in the nucleus.
We also take into account the realistic experimental energy resolution 
 in the results.

First, we show the calculated results at $E_\gamma=2.0$ GeV with the 
potential (\ref{eq:156_29}) in Fig.~\ref{fig:2.0_156_29}. As described
above, this potential is one of 
the
ideal cases to obtain  sharp signals for the mesic states  
formation.  As we can see in Fig.~\ref{fig:2.0_156_29}(a), the peaks due
to the mesic-nucleus formation can be seen clearly in the spectra at
$\theta_p^{\rm Lab}=0$ deg.,  where the momentum transfer  is small
as shown in Fig.~\ref{fig:mom_trans} and the spectra are similar to those
obtained in   
Ref.~\cite{marco} as expected.
Only a limited numbers of subcomponents corresponding to the  
substitutional states are important in this case
as a consequence of the recoilless kinematics. In the spectra,
we can clearly identify the $\omega$ mesic $2p$ state around 
$E_\omega-m_\omega = -50$ MeV.

On the other hand, we found the spectra with significantly different  
shapes at $\theta_p^{\rm Lab}=10.5$ deg. as shown in
Fig.~\ref{fig:2.0_156_29}(b). 
Because of the large momentum transfer around 350 MeV/c at this proton  
angle, many subcomponents have finite contributions to
form the total spectrum, as shown in Fig.~\ref{fig:2.0_156_29}(b), and the
   $\omega$ production spectrum is more similar to a continuum, although only
the excitation of discrete nuclear states is considered in our calculations. 
  The signals of the  
mesic bound states are now smaller than those at 0 degrees.
Thus, it is clear that the experiments at $\theta_p^{\rm Lab}=0$ degrees is  
better suited than those at finite angles
to look for the signals of $\omega$ mesic bound states at $E_\gamma=2.0$ GeV.

In Fig.~\ref{fig:2.0_156_29}, we also show the expected spectra with 
finite   
experimental energy resolutions.  The energy resolution is estimated  
to be around 35-50 MeV in a realistic case \cite{trnkathesis}.  We find in
the figures that the peak structures in the spectrum with the potential
(\ref{eq:156_29})
almost
disappeared  for larger experimental resolutions than  
$\Gamma = 10$ MeV.  Thus, we conclude that an energy resolution better 
than 20 MeV is essentially required to obtain experimental  
evidence of the existence (or non-existence) of the
$\omega$ mesic nucleus.

In Figs.~\ref{fig:2.0_0_50} and \ref{fig:2.0_100_50}, we show the
calculated spectra with potentials (\ref{eq:0_50})
and (\ref{eq:100_50}) at $E_\gamma=2.0$ GeV.  For the potential
(\ref{eq:0_50}) case, we can see   
the enhancement of the cross section at $\theta_p^{\rm Lab}=0$
deg. around $E_\omega - m_\omega = 0$ MeV  in
Fig.~\ref{fig:2.0_0_50}(a).  In this 
case,   
bound states do not exist and the enhancement is due to the
excitation of the $\omega$ to the continuum with recoilless kinematics.  At
$\theta_p^{\rm Lab}=10.5$   
deg., the enhancement is removed by the kinematical conditions with  
the larger momentum transfer as shown in Fig.~\ref{fig:2.0_0_50}(b).  In
Fig.~\ref{fig:2.0_100_50}, the   
spectra with potentital (\ref{eq:100_50}) are shown for $\theta_p^{\rm
Lab}=0$ and $10.5$ 
degrees.  In this case the real part of the optical potential has  
enough attraction to form the bound states in the nucleus, however  
the imaginary part of the optical potential is also strong enough to  
provide a large decay width for these bound states. Thus, we can see  
in the Fig.~\ref{fig:2.0_100_50}(a) that there exists certain strength
under the   
threshold energy which includes the contributions of the bound states  
formation, however,  we cannot identify the binding energies neither  
the widths from the spectra due to the large width of the bound states.   
At $\theta_p^{\rm Lab}=10.5$ deg., we can only see a smooth slope in
the spectra.

We next consider the cases with lower incident energy at  
$E_\gamma=1.2$ GeV, where the momentum transfer at 
$\theta_p^{\rm Lab}=10.5$ deg. takes the smallest value  
 as shown in Fig.~\ref{fig:mom_trans}
Theoretical investigations   
of this kinematics should be important to design  experiments  
which have difficulties for the forward
observation~\cite{trnkathesis}. In   
Fig.~\ref{fig:1.2_156_29} we show the results with the potential
(\ref{eq:156_29}) at this photon   
energy. As shown in Fig.~\ref{fig:mom_trans},  since the momentum
transfer at 0 degree   
is smaller than at 10.5 degree, the signals can be seen clearer at 0  
degree in Fig.~\ref{fig:1.2_156_29}(a) than at 10.5 degree in
Fig.~\ref{fig:1.2_156_29}(b), as expected.   
However, we think it is more important to compare the spectrum in
Fig.~\ref{fig:1.2_156_29}(b)  at 1.2 GeV with that in
Fig.~\ref{fig:2.0_156_29}(b) at 2.0 GeV to 
know the better 
suited incident energy for the observation at finite angles of the 
emitted proton.  We should stress that even if 2.0 GeV allows a 
smaller momentum transfer than 1.2 GeV when the proton is measured 
forward, the signals are clearer  
in the spectrum at 1.2 GeV than at 2.0 GeV at $\theta_p^{\rm Lab}=10.5$ 
degrees, since the momentum  
transfer is smaller for the lower incident photon energy.  In any case, a 
better experimental  
energy resolution than about 20 MeV is required to obtain
decisive information from data on the $\omega$ mesic-nucleus as  
mentioned above.

In Figs.~\ref{fig:1.2_0_50} and \ref{fig:1.2_100_50},  we also show the
calculated spectra with   
potentials (\ref{eq:0_50}) and (\ref{eq:100_50}) at $E_\gamma=1.2$ GeV.
As can be seen in these   
figures, the spectra  show a smooth $\omega$ energy dependence at
this photon energy and the spectra at 0 degree and 10.5 degree resemble  
each other.

As a summary of this section, we would like to add few comments.
In order to obtain the new information on the $\omega$ mesic nucleus  
we need to have the data measured with sufficiently good energy resolution,   
otherwise we can not conclude the existence and/or non-existence  
of the signals due to the mesic-nucleus formation. Furthermore, when planing  
to perform experiments, it is useful to consider the kinematical  
conditions carefully.  In the cases studied here, the incident photon  
energy $E_\gamma=2.0$ GeV is better suited for experiments detecting  
proton emission at $\theta_p^{\rm Lab}=0$ deg, while the lower photon  
energy $E_\gamma=1.2$ GeV is better suited for finite angle proton  
emission at $\theta_p^{\rm Lab}=10.5$ degree.

\section{Monte Carlo simulation of the reaction 
$\gamma A \to \pi^0 \gamma X$}

In this section we perform the Monte Carlo (MC)  computer simulation  of 
the inclusive $A(\gamma,\omega \to \pi^0 \gamma)X$ reaction from 
different nuclei. The method used here (see for the details Ref.~\cite{murat})
combines a phenomenological 
calculation of the intrinsic probabilities for different nuclear
reactions, 
like the quasielastic and absorption channels, as a function of
the nuclear matter density, 
followed by a computer Monte Carlo (MC) simulation procedure in order to 
trace the fate of the $\omega$-mesons and its 
decay products $\pi^0 \gamma$ in the nuclear medium.
Because our MC calculations represent complete event simulations
it will be possible to take into account the actual experimental 
acceptance effects. In the following we shall carry out
the MC simulation taking into account the actual 
geometrical and kinematical acceptance conditions of the 
 TAPS/Crystal Barrel 
experiment at ELSA.

In the MC calculations we shall 
impose the following cuts, both, for the elementary 
$p(\gamma,\omega\to\pi^0\gamma)p$ and photonuclear
$A(\gamma,\omega\to\pi^0\gamma)X$ reactions: 

C\,1) The $\omega$-mesons are produced within an incident beam energy 
constrained by $$1.5~\mbox{GeV}~<~E_{\gamma}^{in}~<~2.6~\mbox{GeV}.$$ 
As in the actual experiment the incident photons are distributed according 
to the unnormalized bremsstrahlung spectrum
\begin{equation}
\label{BRS}
W(E_{\gamma}^{in}) \sim \frac{1}{E_{\gamma}^{in}}.
\end{equation}

C\,2) The polar angle $\theta_p$ of the protons 
produced via a quasi-free kinematics
is required to be detected in the range of
$$7^\circ <~ \theta_p <~ 14^\circ.$$

C\,3) To increase the number of the $\omega \to \pi^0 \gamma$ decays inside
the nucleus the three momentum of the $\pi^0 \gamma$ final state is restricted
to values of $$|\vec{p}_{\pi^0 \gamma}|~
=|\vec{p}_{\pi^0}+\vec{p}_{\gamma}|~<~400~\mbox{MeV}.$$
Indeed, the fraction of the $\omega$ mesons decaying inside the nucleus can be
optimized by minimizing the decay length
$L_\omega = (p_\omega/m_\omega\Gamma_\omega)$ where $\Gamma_\omega$
is the width of the $\omega$ in the rest frame.
It is therefore preferred that the kinetic energy of the $\omega$-mesons 
reconstructed from $\pi^0 \gamma$ events 
with three momenta $\vec{p}_{\omega} = \vec{p}_{\pi^0}+\vec{p}_{\gamma}$
is small.

C\,4) The kinetic energy  $T_{\pi^0}=E_{\pi^0}-m_{\pi^0}$ 
of the detected $\pi^0$ 
is taken to be larger
   than 150 MeV. This cut will strongly suppress the distorted events
due to the quasielastic $\pi^0$ final state interactions (FSI) 
with the nucleons
of the target.

C\,5) The energy of the photon in the $\pi^0 \gamma$ final state is larger than
   $200$~MeV. This cut should attenuate the background channels.


We start our MC analysis with the cross section of 
the elementary reaction $\gamma p \to \omega p \to \pi^0\gamma p$.
In Ref.~\cite{EPJA18} the total cross section and 
the invariant differential cross sections 
$d\sigma_{\gamma p \to \omega p}/dt$ 
of the reaction $(\gamma,\omega)$ on protons were
measured
for incident photon energies from the reaction
threshold $E_{\gamma}^{th} = m_{\omega}+m_{\omega}^2/2M_p \simeq 1.1$~GeV 
up to 2.6~GeV. We use this experimental information in our analysis 
which is conveniently parametrized in Ref.~\cite{murat}. In the following, 
the cross section on the neutron will be assumed to be the same as 
on a proton.

\begin{figure*}[t]
\begin{center}
\includegraphics[clip=true,width=0.74\columnwidth,angle=0.]
{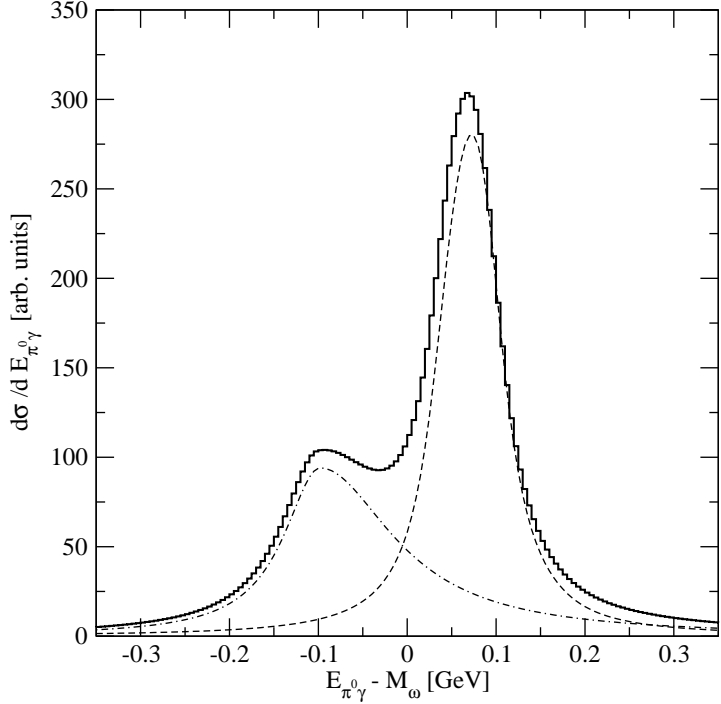}
\caption{\label{Fig2} \small
The differential cross section $d\sigma/dE_{\pi^0 \gamma}$ 
of the reaction $\gamma p \to \pi^0 \gamma p$ as a function of the
$E_{\pi^0\gamma}-m_{\omega}$ where $E_{\pi^0\gamma}=E_{\pi^0}+E_{\gamma}$.
 The spectrum (solid histogram) 
is obtained using the reconstructed $\pi^0\gamma$
events from the exclusive
$\gamma p \to \omega(\pi^0\gamma) p$ reaction (dashed curve) 
plus an inclusive  $\pi^0 \gamma$ background discussed in the text (dash-dotted curve). 
The following cuts were imposed: 
$E_{\gamma}^{in}= 1.5 \div 2.6$~GeV, $7^{\circ} < \theta_p < 14^{\circ}$,
$|\vec{p}_{\pi^0}+\vec{p}_{\gamma}| < 400$~MeV, $|\vec{p}_{\gamma}| > 200$~MeV and
$T_{\pi^0} > 150$~MeV. The exclusive $\omega \to \pi^0 \gamma$ signal has been
folded with the 50 MeV experimental resolution.}
\end{center}
\end{figure*}

Our results for the differential 
cross section $d\sigma/dE_{\pi^0 \gamma}$  as a function
of the $E_{\pi^0 \gamma} - m_{\omega}$ 
where $E_{\pi^0 \gamma} = E_{\pi^0} + E_{\gamma}$, 
and after applying the experimental cuts listed above,
are shown in Fig.~\ref{Fig2}. 
There are preliminary data for this reaction from the CBELSA/TAPS experiment. The lack
of definitive data to which we could compare our results should not be an
obstacle to discuss the theoretical results and draw some conclusions. Apart
from the cross section from $\gamma p \to \omega p \to \pi^0 \gamma p$ that we
evaluate, there should be background events from $\pi^0 \gamma p $ events where
the $\pi^0 \gamma$ does not come from the decay of the $\omega$. These and other
background events are certainly present in the reaction, as shown in 
\cite{trnka}, and they have larger strength at invariant masses lower
than $m_{\omega}$. Correspondingly, there should be some background in the
$\pi^0 \gamma$ energy distribution at lower energies than $m_{\omega}$. We have
evaluated the phase space for the $\gamma p \to p \pi^0 \gamma $  reaction and 
find indeed strength  at $\pi^0 \gamma$ energies below $m_{\omega}$.    
There can also be other sources of background like from $\gamma p \to \pi^0
\pi^0 p$, or $\gamma p \to \pi^0 \eta p$, where one of the two photons from
the decay of the $\pi^0$ or the $\eta$ is not measured.
We do not want to make a theory of the background
here, but simply justify that a background like the one assumed in  Fig.~\ref{Fig2},
peaking around $m_{\omega} -100$~MeV is rather plausible.  Indeed, this seems to be
the case from the preliminary data of CBELSA/TAPS, with a background very similar to that drawn in
Fig.~\ref{Fig2}. Yet, the conclusions of this section are not tied to specific details
of this background but to general features which we discuss below. 
   In Fig.~\ref{Fig2} the solid histogram
is obtained as a sum of the reconstructed 
exclusive events from the 
$\gamma p \to \omega p \to \pi^0\gamma p$ reaction (dashed curve) 
and the background contribution. 
For the exclusive $\pi^0 \gamma$ 
events coming from $\gamma p \to \omega p \to \pi^0\gamma p$ 
an experimental resolution of $50$~MeV was imposed, 
see Ref.~\cite{trnka}. Note that the last two cuts in the list, 
$E_{\gamma} > 200$~MeV
and $T_{\pi^0} > 150$~MeV, are irrelevant for the $\omega \to \pi^0 \gamma$
events 
since basically all of the MC events coming from this source 
fall in the kinematic regions allowed by these cuts.

In the photonuclear reaction $A(\gamma,\omega)X$
the $\omega$-mesons are produced inside the nucleus according to their 
in-medium spectral function which includes
the collisional broadening of the $\omega$ due to the quasielastic and
absorption channels.
For the quasielastic scattering of the $\omega$ we use the parametrisation of 
the $\omega N \to \omega N$ cross section given in 
Refs.~\cite{lykasov,muhlich2}. The in-medium 
quasielastic scattering does not lead to a loss of flux
and therefore does not change 
the total nuclear cross
section.  But the later affects the $\omega$ momentum and energy 
distributions, 
keeping the $\omega$ meson inside the nucleus for a longer time. 
The loss of $\omega$ flux is related to the absorptive part of the 
$\omega$-nucleus optical
potential.
In the following the absorption width of the $\omega$ will be taken in the 
form 
\begin{equation}
\label{Gabs}
\Gamma_{abs} \simeq 100 \, \mbox{MeV} \times \frac{\rho(r)}{\rho_0}
\end{equation}
with no shift of the $\omega$ mass, as suggested by the analysis
of~\cite{murat},
although this latter assumption is not relevant for the conclusions to be drawn.
As will be shown in Ref.~ \cite{cola} ( preliminary results are available in 
\cite{trnkathesis}), the in-medium 
$\omega$ width of $\simeq 100$~MeV at normal nuclear 
matter density explains the
nuclear transparency ratio measured in Ref.~\cite{trnkathesis}.

Using the MC method of Ref.~\cite{murat}, which 
proceeds in a close analogy to the actual
experiment, we 
trace
the fate of the $\omega$-mesons and their decay products
in their way out of the nucleus. All standard nuclear effects like the
Fermi motion of the initial nucleons and Pauli blocking of the final ones
are taken into account. 
The $\omega$-mesons are allowed to propagate inside the nucleus and
at each step $\delta L$ the reaction probabilities for different 
channels like
the decay of the $\omega$
into $\pi^0 \gamma$ and $\pi\pi\pi$ final states, 
quasielastic scattering and in-medium absorption
are properly calculated. Since we are interested in $\pi^0\gamma$ events
the absorption channels and decay $\omega \to \pi\pi\pi$ 
remove the $\omega$-mesons from initial flux.
The reconstructed $\pi^0 \gamma$ events may come from both the $\omega$
decaying inside and outside the nucleus. 
If the resonance leaves the nucleus, its spectral function must coincide with
the free distribution, 
because the collisional part of the width is zero in
this case.
When the $\omega$ decays into $\pi^0 \gamma$ pair inside the nucleus 
its mass distribution is generated according to the 
in-medium spectral function at the local density. 
For a given mass $\widetilde{m}_{\omega}$ 
the $\omega$-mesons are allowed to decay 
isotropically in the c.m. system 
into $\pi^0 \gamma$ channel
with a width
$\Gamma_{\pi^0\gamma}= 0.76$~MeV.
 The direction of the $\pi^0$
(therefore $\gamma$)
is then chosen randomly  and an appropriate Lorentz transformation is done
in order to generate the corresponding $\pi^0 \gamma$ distributions in the
laboratory frame. 
 The
$\omega$-mesons are reconstructed using the energy and 
momentum of the $\pi^0 \gamma$
pair in the laboratory.

\begin{figure*}[t]
\begin{center}
\includegraphics[clip=true,width=0.75\columnwidth,angle=0.]
{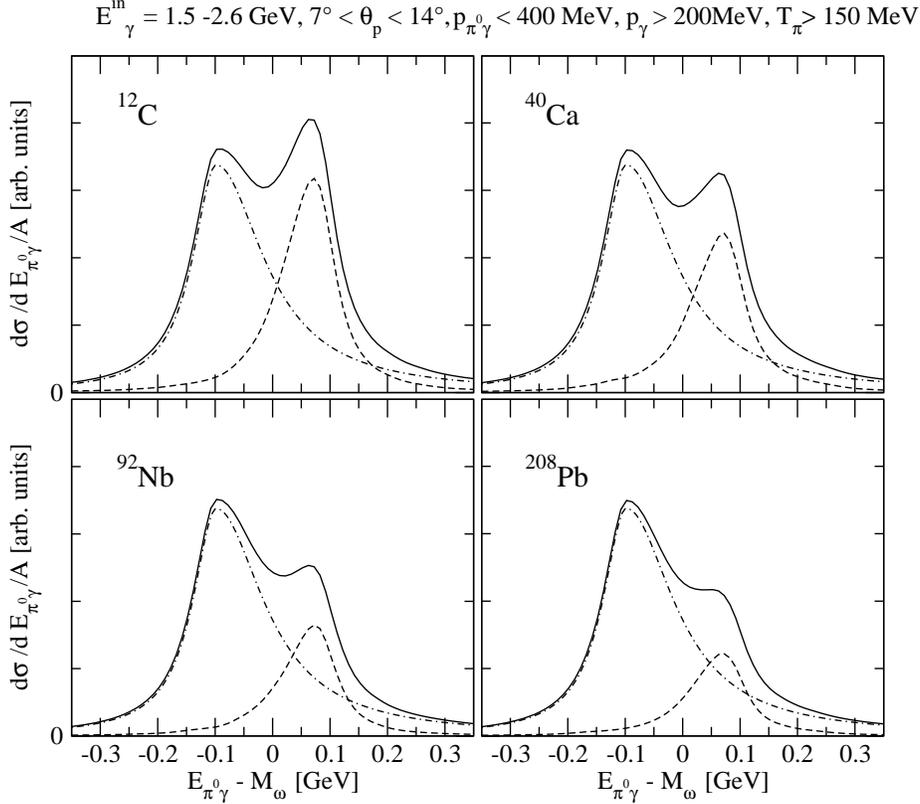}
\caption{\label{FigEMnuclear} \small
The differential cross section $d\sigma/dE_{\pi^0\gamma}$ 
of the reaction $A(\gamma,\pi^0\gamma)X$
as a function
of $E_{\pi^0\gamma} - m_{\omega}$  from $^{12}$C, $^{40}$Ca, $^{92}$Nb and
$^{208}$Pb nuclear targets. The reconstructed exclusive events from the 
$\omega \to \pi^0 \gamma$ decay are
shown by the dashed curves.  The $\pi^0 \gamma$ background is shown by
the dash-dotted curves. 
The sum of the two contributions is given by the solid curves.
The following cuts were imposed: 
$E_{\gamma}^{in}= 1.5 \div 2.6$~GeV, $7^{\circ} < \theta_p < 14^{\circ}$,
$|\vec{p}_{\pi^0}+\vec{p}_{\gamma}| < 400$~MeV, $|\vec{p}_{\gamma}| > 200$~MeV 
and $T_{\pi} > 150$~MeV. The exclusive $\omega \to \pi^0 \gamma$ signal 
has been folded with the 50 MeV experimental resolution.
All spectra are normalized to the corresponding 
nuclear mass numbers $A$.} 
\end{center}
\end{figure*}

\begin{figure*}[t]
\begin{center}
\includegraphics[clip=true,width=0.75\columnwidth,angle=0.]
{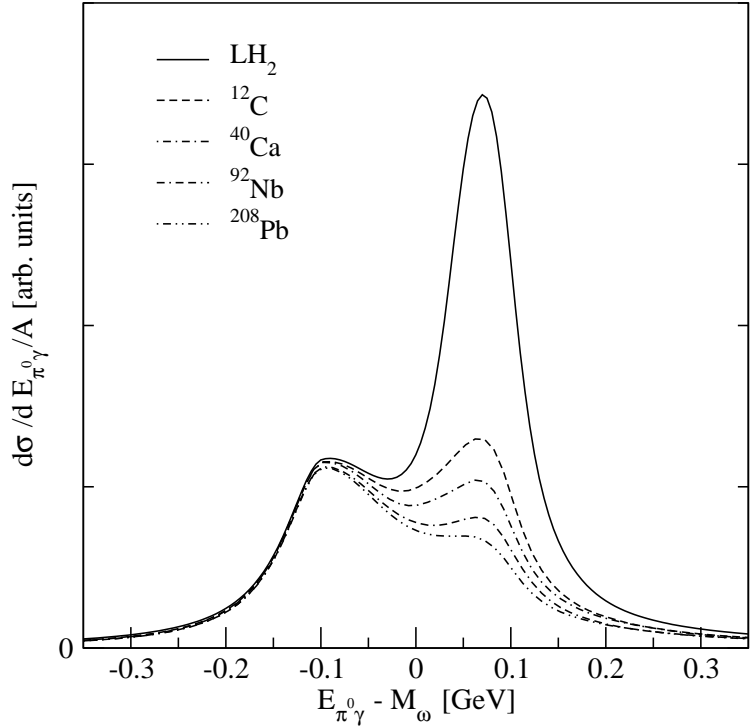}
\caption{\label{Fig4fin} \small
Summary plot of the reconstructed $\pi^0 \gamma$ events 
in the reaction $(\gamma,\pi^0\gamma)$ from the
proton target (solid) 
and sample nuclear targets $^{12}\mbox{C}$ (dashed), 
$^{40}\mbox{Ca}$(dash-dotted), 
$^{92}\mbox{Nb}$(dash-dash-dotted) and $^{208}\mbox{Pb}$(dot-dot-dashed).} 
\end{center}
\end{figure*}

The reconstruction of the genuine
$\omega{\to}\pi^0\gamma$ mode is affected by the final state
interactions of the $\pi^0$  in the nucleus. 
In this case, if the $\pi^0$ events come from the interior of the nucleus 
we trace the fate of the neutral pions 
starting  
from the decay point of the $\omega$-meson.
 In their way out
of the nucleus pions can experience the quasielastic scattering or 
can be absorbed. The intrinsic probabilities for these reactions
as a function of the nuclear matter density
are calculated using the phenomenological models of 
Refs~\cite{Salcedo:1987md,Oset:1986sy,Oset:1990zj},
which also include 
higher order quasielastic cuts and 
the two-body and three-body
absorption mechanisms.
Note that the FSI of the pions distorts the
$\pi^0 \gamma$ spectra and makes also an additional 
contribution to the negative part of the $E_{\pi^0 \gamma} - m_{\omega}$
distribution. It was already demonstrated in 
Refs.~\cite{Messchendorp:2001pa,muhlich} that the contributions
of the distorted events due to the FSI of the pions
can be largely suppressed by using the cut on kinetic energy of pions
$T_{\pi} > 150$~MeV. 
We confirm this finding and use it further in our analysis. 
Since the FSI of the $\gamma$ quanta 
are rather weak they  are allowed to escape the nucleus without distortion.

\section{Results of the MC calculations}

In the following we assume that the inclusive
$\pi^0 \gamma$ background scales with respect to 
the target nucleus 
mass number $A$ like
\begin{equation}
\label{scale}
\sigma_A \simeq A \, \sigma_{elem}.
\end{equation}
This assumption  implies merely a rather weak absorption of 
the inclusive $\pi^0 \gamma$ pairs
in the nuclear medium. 
Note that this assumption is supported by the present MC calculations which 
show only the marginal effects due to the FSI of the relatively fast pions
(beyond the $\Delta(1232)$ region). 
Recall that, because of the cuts we have imposed, the kinetic energies of the
pions are always larger than $T_{\pi} > 150$~MeV.
But this is not
the case for the exclusive $\pi^0 \gamma$ events coming from the decay of the
$\omega \to \pi^0 \gamma$. Although, the pions coming from this source
are also fast and easily abandon the nucleus without significant FSI, 
the rather strong absorption
of the $\omega$ inside the nucleus might change the scaling relation
Eq.~(\ref{scale}) and
\begin{equation}
\sigma_A(\omega \to \pi^0 \gamma) \simeq A^{\alpha} \, 
\sigma_{elem}(\omega \to \pi^0 \gamma)
\end{equation} 
where the attenuation parameter $\alpha < 1$.

In Fig.~\ref{FigEMnuclear} 
we show the result of the MC simulation for the 
$E_{\pi^0 \gamma} - m_{\omega}$ 
spectra reconstructed from the $\pi^0$ and $\gamma$ events. The calculations
are performed 
for the sample nuclear targets $^{12}\mbox{C}$, $^{40}\mbox{Ca}$, 
$^{92}\mbox{Nb}$ and $^{208}\mbox{Pb}$. The kinematic and acceptance
cuts discussed before  have been already imposed. 
The MC distributions are
normalized to the nuclear mass number $A$. The solid curves correspond
to the sum of the inclusive $\pi^0 \gamma$ background (dash-dotted curve), 
which is not related to the
 production and decay of the $\omega$-mesons, and the exclusive $\pi^0 \gamma$ 
events coming from the direct
decay of the $\omega\to \pi^0 \gamma$. The contributions of the exclusive
$\omega \to \pi^0 \gamma$
events are shown by the dashed curves. We note a
very strong attenuation of the
$\omega\to \pi^0 \gamma$ signal with respect to the background contribution
 with increasing  nuclear mass number $A$.
 This is
primary due to the stronger absorption of the $\omega$-mesons 
with increasing nuclear matter density, see Eq.~(\ref{Gabs}).
Also the contribution of 
the $\omega$-mesons decaying inside the nucleus (with and without 
$\pi^0$ rescattering) is increasing as a function of mass number 
$A$ merely due to an increase in the effective radius of the nucleus.
In Fig.~\ref{Fig4fin} we compare the nuclear cross sections 
of all sample nuclei with the 
$E_{\pi^0 \gamma} - m_{\omega}$ spectra from the hydrogen
target. Here one can see a double hump structure
of the $E_{\pi^0 \gamma} - m_{\omega}$ spectra  but with the considerable
attenuation of the exclusive $\omega \to \pi^0\gamma$ signal 
from light to heavy nuclei. 

  The former excercise indicates that given the particular combination of 
$\pi^0 \gamma$ from an uncorrelated background and from  $\omega$ decay, and
the different behaviour of these two sources in the $\pi^0 \gamma$ production in
nuclei, a two bump structure seems unavoidable in nuclei. Should we have not
done this calculation and intepreted it in the way we have done, the observation
of a peak at about 100 MeV below the $\omega$ mass could easily tempt anyone to
claim it as an indication of a bound $\omega$ state in the nucleus. By
performing the present study we are paving the way for future investigations on
the topic with the performance of the necessary simulation of conventional
mechanisms which accompany the reactions used in the search for more exotic 
phenomena. 

   We would like to insist on the issue of the background, because there can be many
   sources for it. In the CBELSA/TAPS experiment, where the $\pi^0$ is observed from the two
   $\gamma$ decay, a background of  $\pi^0 \gamma $ could also come from the
   production of two $\pi^0$, or $\pi^0 \eta$, with no detection of a fourth
   $\gamma$
   coming from one $\pi^0$ or $\eta$ decay.  Such background
   could be eliminated, but there are unavoidable backgrounds like that coming from
   the $\gamma p \to p \pi^0 \gamma$  reaction.
   Having this in mind we can easily see that even if the background of the reaction
   after eliminating avoidable backgrounds is about one third of the one assumed in the
present discussion, or even smaller, the two bump structure of  Fig.~\ref{FigEMnuclear}
 would still show up.
 
  Finally, we would like to make a final comment in the sense that the strength of the
  cross section bears some information in itself.  In the calculations done for the
  capture of $\omega$ mesons in bound orbits the cross sections are presented in
  absolute numbers. Should  experiments  find some peak structure in
  the $\omega$ bound region with a strength considerably larger than the one predicted
  in the present calculations, this would be an indication that the strength observed
  is coming from some sort of background, not from the formation of bound $\omega$
  states.  


\section{Conclusions}
  In the present work we have carried out some calculations which should be very
helpful in the search for eventual bound $\omega$ states in nuclei. In the first
part we evaluated the reaction cross sections for the $(\gamma , p) $ reaction
in nuclei leading to the production of bound or unbound $\omega$ states together
with the  excitation of nuclear bound states.   The calculations were done using
different optical potentials which covered a wide range of bindings and 
absorptive parts.  We found that only for potentials where the real part was 
larger
than twice the imaginary part there was some chance to see peaks in the $\omega$
energy spectrum corresponding to the formation of the $\omega$ bound states. 
Clear peaks could be seen for a potential (-156, -29) MeV ( at $\rho = \rho
_0$), while if we had an imaginary part of about -50 MeV, as suggested by present
experimental data on the A dependence of the $\omega$ production,  even with 100
MeV binding no signal could be seen in the calculated spectrum. We  studied
the reaction for different photon energies and different proton angles. Since
the optimal situation to see the peaks appears for recoilesss kinematics, the
photon energy of 2 GeV was the optimal one if one observes protons in the
forward direction.  However, if the experimental conditions make it impossible
or difficult to measure forward protons  and a proton angle around 10 degrees 
is the choice, then we showed that a photon energy of about 1.2 GeV was more
suited and led to better recoilles kinematics than with photons of 2 GeV.  We
performed the calculations for this situation which should serve to compare with
experimental measurements made in the future.

Another relevant finding of the present work was that, even if bound states
exist and they lead to peaks in the $(\gamma , p) $ reaction, an experimental
resolution better than 20 MeV in the $\omega$ energy is mandatory to make the
peaks visible. 

   Finally we made another useful evaluation by calculating the inclusive
  $(\gamma , p) $ reaction leading to $\pi ^0 \gamma$ events. To the elementary
  reaction $\gamma p \to \omega p$
   with subsequent $\pi ^0 \gamma$ decay of the $\omega$ 
   we added a certain
   background  from reactions leading to $\pi ^0 \gamma$ with not connection to 
   $\omega$ production. Then we studied 
  this reaction in nuclei, and because of the different behavior in the
  production of uncorrelated and $\omega$ correlated $\pi ^0 \gamma$ pairs, we
  showed that a peak appeared in nuclei in the region of $\pi ^0 \gamma$ energy
  around $m_{\omega}-100$~MeV, which could easily be misidentified by a signal of
  a bound $\omega $ state in nuclei.  
  
  Altogether, the information found here should be of much help in order to
  identify the ideal set ups for future experiments searching for 
  $\omega$ bound states in nuclei and to properly interprete  results of these
  experiments.

\section*{Acknowledgments}
We would like to acknowledge useful discussions with V. Metag, 
M. Kotulla and D. Trnka.
This work is partly supported by DGICYT contract number BFM2003-00856,
and the E.U. EURIDICE network contract no. HPRN-CT-2002-00311.
This research is part of the EU Integrated Infrastructure Initiative
Hadron Physics Project under contract number RII3-CT-2004-506078.
One of the authors (H.N.) is supported by Research Fellowship of JSPS  
for Young Scientist.  This work is partly
supported by Grants-in-Aid for scientific research of JSPS 
(No.  16540254 (S.H.) and No. 18-8661 (H.N.)),
and partly supported by CSIC and JSPS under the Spain-Japan 
Research  Cooperative Program.

\end{document}